# Role of thermal vibrations in magnetic phase transitions

T. R. S. Prasanna

Department of Metallurgical Engineering and Materials Science

Indian Institute of Technology, Bombay

Mumbai – 400076

India

We address and resolve the fundamental contradiction that has existed from the earliest studies on magnetic phase transitions between theoretical models that ignore the role of thermal vibrations and represent the exchange interaction as a constant, $J_{ij}(0)$, and analysis of neutron diffraction data that always incorporates thermal vibrations even though it is also possible to analyze the same data by ignoring them. Of the two possibilities, ignoring thermal vibrations in both theoretical models and analysis of diffraction data leads to the latter giving different magnetic order parameters for different reciprocal lattice lines. *This appears to be the first report of a unique consequence, viz. the assumption to neglect a physical phenomenon turns a single-valued experimental observable into a multiple-valued one where all values are equally valid.* This assumption is clearly unacceptable and must be rejected. The second possibility of incorporating thermal vibrations in both leads to single-valued theoretical and experimental order parameters. Thus, analysis of neutron diffraction data constrain the exchange interaction in all theoretical models to be temperature dependent and represented as $J_{ij}(T)$. Additional experimental and theoretical evidences in support of this conclusion are presented.

The Heisenberg model is considered to be the prototype model for magnetic phase transitions and is represented as [1-3]

$$H(0) = -\sum J_{ij}(0) \, \boldsymbol{S_i}.\boldsymbol{S_j} \qquad (1)$$

However, the exchange interaction parameter, $J_{ij}(0)$, in Eq.1 is for a static lattice (T = 0 K). As is well known, the exchange interaction parameter is a function of the wavefunctions of at least two electrons, *i and j,* whose spins are interacting. It can be a function of more that the wavefunctions of the said two electrons in case of indirect or superexchange mechanisms. Therefore, in all cases, the exchange interaction can be symbolically represented for a static lattice (T = 0 K) as

$$J_{ij}(0) = J_{ij}\big(\psi_i(0),\, \psi_k(0),\, \psi_l(0) \dots \psi_j(0)\big) \qquad (2)$$

where $\psi_k(0), \psi_l(0)$ are present only if the exchange interaction between electrons *i and j* is mediated by other ions. *The implicit assumption in Eq.1 is that the role of thermal vibrations in phase transitions is negligible i.e. whether the role of thermal vibrations is incorporated or ignored makes little difference to theoretical or experimental results.*

As is well known, magnetic neutron diffraction is the primary technique to determine the magnetic structure and the state of order, especially for antiferromagnetic substances. From the very beginning till date, analysis of neutron diffraction data always incorporates the role of thermal vibrations [4-6] *even though it is also possible to analyse the same data by ignoring thermal vibrations*. This contradiction between theoretical models and analysis of neutron diffraction data has gone unnoticed till date.

Recently, we have addressed [7] a similar contradiction in order-disorder alloy phase transitions and shown that diffraction data impose severe constraints on all theoretical models. It is essential to incorporate thermal vibrations for a correct understanding of order-disorder transitions. In particular, all existing models must be modified to explicitly include a temperature dependent interaction parameter before their predictions can be compared with diffraction data [7].

While the main ideas in alloy and magnetic phase transitions are similar, some differences exist. Firstly, in alloy phase transitions, the ordered and disordered phases can have different elastic properties leading to a vibrational entropy contribution. This will be absent in magnetic phase transitions since the chemical environment remains the same. Secondly, in alloy phase transitions critical temperatures are ~ 1000 K which implies that the constraint imposed by diffraction data is severe. However, in magnetic phase transitions, the transition temperatures vary from 1 K – 1000 K. Hence, the role of thermal vibrations must be examined more closely.

In this paper, we address and resolve the contradiction between theoretical models of magnetic phase transitions that ignore thermal vibrations and analysis of magnetic neutron diffraction data that always incorporates the role of thermal vibrations.

To explain the temperature dependence of band structures, especially in semiconductors, finite temperature band structure theory was developed [8-10]. This formalism incorporates the role of thermal vibrations in electronic structure theory and the correct approach has been described [8] as "*A higher order adiabatic perturbation summation can be accomplished by solving $H_0 + \overline{H_2} + \overline{H_4} + ...$ exactly (Keffer et. al. 1968) and then using the resulting temperature-dependent eigenfunctions and energies to calculate the self-energy terms.*" That is, the first step is to solve the electronic structure for a temperature dependent core potential given by

$$V_i(\boldsymbol{G}, T) = V_i(\boldsymbol{G}, 0) e^{-W_i(\boldsymbol{G}, T)} \qquad (3)$$

where $V_i(\boldsymbol{G}, 0)$ is the static lattice (0 K) core potential of ion "$i$" and $W_i(\boldsymbol{G}, T)$ is the Debye-Waller factor (DWF). Even before the formal development of the finite temperature band structure theory [8-10], this form of the core potential was empirically used by Keffer [11] and others [12,13] to explain the temperature dependence of band structures. This formalism has been used to explain several high temperature valence electron properties in metals and semiconductors [11-15].

It is clear that the high temperature electronic structure theory results in "*temperature-dependent eigenfunctions and energies*" i.e. $\psi(T)$ and $\varepsilon_{n\boldsymbol{k}}(T)$ respectively. It follows that, at finite temperatures, the exchange interaction is a function of temperature dependent wavefunctions and must be represented as

$$J_{ij}(T) = J_{ij}\big(\psi_i(T), \psi_k(T), \psi_l(T) \ldots \psi_j(T)\big) \qquad (4)$$

Thus, at finite temperatures, the prototype Heisenberg model is modified from Eq.1 as

$$H(T) = -\sum J_{ij}(T)\, \mathbf{S}_i \cdot \mathbf{S}_j \qquad (5)$$

The high temperature electronic structure theory described above [8-10] is based on the adiabatic approximation and is valid for insulators and semiconductors at all temperatures and for metals above the Deybe temperature ($\Theta_D$). Thus for insulators and semiconductors at all temperatures and for metals above $\Theta_D$, *Eq.5 is the correct representation of the prototype Heisenberg model.* Eq.1 is an approximation to Eq.5 that ignores the role of thermal vibrations.

In unpolarized neutron diffraction, the magnetic intensity is proportional to the magnitude square of the magnetic structure factor, $\left|F_G^{mag}\right|^2$ which is given by (Eq.6.14 of Ref.5)

$$\left|F_G^{mag}\right|^2 = \left|\sum p\, e^{2\pi i \mathbf{G}\cdot \mathbf{r}_j}\right|^2 e^{-2W(\mathbf{G},T)} \qquad (6)$$

where W($\mathbf{G}$,$T$) is the Debye-Waller factor and $p$ is the magnetic scattering amplitude that is related to the magnetic form factor, $f$, by the relation [Eq.6.10 of Ref.5]

$$p = \left(\frac{e^2 \gamma}{mc^2}\right) Sf \qquad (7)$$

As is well known [16], the magnetic order parameter, magnetization or sublattice magnetization, at any temperature is proportional to the magnetic intensity. It is obtained from the relation

$$I_G(T) = \eta^2(T)\, I_G(0)\, e^{-2W(G,T)} \qquad (8)$$

where $\eta(T)$ is the magnetic order parameter, $I_G(0)$ is the magnetic intensity of line $G$ when the system is fully ordered ($\eta(0) = 1$) under static lattice (0 K) conditions.

Comparing Eq.3 and Eq.6, both the high temperature electronic structure theory and neutron diffraction theory incorporate the role of thermal vibrations through the Debye-Waller factor. Clearly, if theoretical models ignore the role of thermal vibrations then the same assumption must be applied to the analysis of experimental neutron diffraction data as well. That is, *the DWF must be simultaneously incorporated or ignored (W=0), both in Eq.3 and Eq.6, Eq.8.*

In most studies, the magnetic intensity given by Eq.8 is plotted as a function of temperature, though infrequently, the magnetic order parameter (sublattice magnetization) given by Eq.9a (below) has been determined as a function of temperature [17]. The magnetic order parameter is obtained from Eq.8 as

$$\eta(T) = \left[I_G(T)/\, I_G(0)\, e^{-2W(G,T)}\right]^{1/2} \qquad (9a)$$

The magnetic order parameter obtained from Eq.9a must be compared with predictions from models based on Eq.5 since both incorporate thermal vibrations. However, if the role of thermal vibrations is ignored by setting *W=0*, the magnetic order parameter is given by

$$\eta(T) = [I_G(T)/I_G(0)]^{1/2} \tag{9b}$$

The magnetic order parameter obtained from Eq.9b must be compared with predictions from models based on Eq.1 since both ignore thermal vibrations. However, in Eq.9b, different lines *G* give different values of the magnetic order parameter.

From the above discussion it is clear that the current representation of the magnetic order parameter as $\eta(T)$ is inadequate. We propose a new representation that reflects its dependence on several variables. The experimental magnetic order parameter depends on the reciprocal lattice line *G* and also on whether thermal vibrations are incorporated ($W=W_D$) or ignored ($W=0$). Thus, the experimental magnetic order parameter obtained from Eq.9a must be represented as $\eta_G^{exp}(T/T_c, W_D)$ and that obtained from Eq.9b must be represented as $\eta_G^{exp}(T/T_c, 0)$. The theoretical magnetic order parameter has no dependence on the reciprocal lattice vector, *G*, but depends on whether thermal vibrations are incorporated ($W=W_D$) or ignored ($W=0$). Thus, the theoretical magnetic order parameter obtained from Eq.5 must be represented as $\eta^{th}(T/T_c, W_D)$ and that obtained from Eq.1 must be represented as $\eta^{th}(T/T_c, 0)$.

The above results are summarized in Table 1. It is clearly seen that comparisons between theoretical and experimental magnetic order parameters can only be between Eq.5 – Eq.9a and Eq.1 – Eq.9b, since they would be under the same assumption. In Eq.9a, all reciprocal lattice lines *G* give the same (single) value of the magnetic order parameter that must be compared with the single value predicted from Eq.5. In Eq.9b, different lines $G$ will give different values of the magnetic order parameter which must be compared with a single value predicted from Eq.1. This is clearly unacceptable and is further discussed later. The current practice of comparing the predicted magnetic order parameter from Eq.1 with the experimental magnetic order parameter obtained from Eq.9a (as both give single-valued order parameter) is incorrect as they have been obtained under different assumptions and is clearly evident from Table 1.

However, since current models are based on Eq.1, it is clear that $\eta^{th}(T/T_c, 0)$ must be compared only with $\eta_G^{exp}(T/T_c, 0)$. We present a simple method that involves minimum disruption of the current practice of comparing the predictions of Eq.1 with $\eta_G^{exp}(T/T_c, W_D)$. If the experimentally determined error due to neglect of thermal vibrations is reported along with $\eta_G^{exp}(T/T_c, W_D)$, this information is *equivalent to comparing* $\eta_G^{exp}(T/T_c, 0)$ *with* $\eta^{th}(T/T_c, 0)$.

The percentage error due to the neglect of thermal vibrations is readily seen to be

$$\varepsilon_{vib}^{\%}(\boldsymbol{G}, T) = \left(1 - \frac{\eta_{\boldsymbol{G}}^{exp}(T/T_c, 0)}{\eta_{\boldsymbol{G}}^{exp}(T/T_c, W_D)}\right) \times 100 = \left(1 - e^{-W(\boldsymbol{G}, T)}\right) \times 100 \quad (10)$$

The error in Eq.10 is *always positive* and neglecting the role of thermal vibrations (Eq.9b) leads to magnetic order parameter that is always lower than that obtained by incorporating them (Eq.9a). This error subtracted from the correct value of the magnetic order parameter, $\eta_G^{exp}(T/T_c, W_D)$, gives $\eta_G^{exp}(T/T_c, 0)$ which must be compared with $\eta^{th}(T/T_c, 0)$. *Therefore, as a matter* *of scientific correctness this error must always be reported in all experimental analysis of neutron diffraction data since it reflects that both theoretical models and experimental analysis have ignored thermal vibrations.* It is stressed that this comparison is still scientifically unacceptable since different reciprocal lines $G$ give different magnetic order parameter values.

From the second equality in Eq.10, it is clear that it is unnecessary to know the absolute value of either of two magnetic order parameters in order to determine the percentage error. In addition, the accurate determination of $W(G,T)$ by *any diffraction technique* can be used to estimate the percentage error in the magnetic order parameter when thermal vibrations are ignored. The advantage of this feature will become clearer when specific systems are discussed below. Since $W(G,T)$ is always determined during the analysis of neutron diffraction data, the error can be readily determined without performing any new experiments. As is well known [2,3,18], the DWF, $W(G,T)$, is given by

$$W = 8\pi^2 \langle u_s^2 \rangle \left(\frac{\sin\theta}{\lambda}\right)^2 = B(T)\left(\frac{\sin\theta}{\lambda}\right)^2 \qquad (11)$$

where $\langle u_s^2 \rangle$ is the mean-squared displacement and $B(T)$ is the isotropic temperature factor. From Eq.10 and Eq.11 it is clear that the error depends on the reciprocal lattice line $G$ and the temperature of interest. The error will be largest near the critical temperature for any given reciprocal lattice line $G$ and for a fixed temperature will increase with increasing $|G|^2$. Since, the first reciprocal lattice line occurs at a finite value of $|G|^2$, a *minimum* error due to the neglect of thermal vibrations will always be present. Since the mean-squared displacement, $\langle u_s^2 \rangle$, or the isotropic temperature factor, B($T$), increase with temperature, the error in magnetic systems with higher transition temperatures will be higher than in systems with lower transition temperatures.

As is well known [4,19-23], the transition metal oxides, MnO, CoO and NiO are prototype systems to study anti-ferromagnetism. These systems have been well studied and accurate diffraction data are available for these systems. We calculate the errors for these oxides when their experimental magnetic order parameters have to be compared with predictions made from static lattice prototype Heisenberg models based on Eq.1. We use the data from Ref.20-22 as the value of the mean-square amplitude at room temperature for CoO in Ref.20 ($U_{Co}$ = 0.00518 Å$^2$) is also independently confirmed in a polarized neutron diffraction study ($U_{Co}$ = 0.0050 Å$^2$) by a different group [23]. In Ref.20-22, the data are reported for MnO at 15 K and 295 K (far away from $T_N$), for CoO at 10 K and 305 K and for NiO at 10 K and 550 K. For CoO and NiO, the high temperature data are just above the Neel temperature of 293 K and 523 K respectively. Using the well known assumption that at high temperatures, DWF varies linearly with temperature, a 5% reduction is made in the mean-square displacement values reported in Ref.20 and Ref.22 so that they represent the displacements very close to the Neel temperature. Thus, $U_{Co}$ = 0.00492 Å$^2$ (~ $T_N$) and $U_{Ni}$ = 0.00594 Å$^2$ (~ $T_N$) are used. Table 2a summarizes the error due to the neglect of thermal vibrations in MnO, CoO and NiO at very low temperatures and near $T_N$. They vary from 0.5% - 4.5% depending on the temperature and the superlattice line ***G***.

Table 2b summarizes the errors due to the neglect of thermal vibrations in ferrimagnetic NiFe$_2$O$_4$ with a $T_N$ ~ 860 K. The data is taken from Ref.24 and Ref.25 and $B_{iso}$(850 K) is obtained from linear extrapolation of $B_{iso}$(300 K) in Ref.25. The reciprocal lattice lines have been chosen to have majority contribution from magnetic scattering and in particular for (331) the entire intensity is due to magnetic scattering [24]. The errors due to neglect of thermal vibrations are unacceptably large near $T_N$. In particular, for systems where $T_N$ is high, the error analysis shows that the assumption of constant exchange integral is likely to be incorrect. Hence for spinel ferrites where $T_N$ is high (~ 800 K), the long standing practice [26] of extracting constant exchange interaction parameters from experimental data using mean-field models must be modified to have temperature dependent exchange interactions.

All superlattice lines give a single-valued order parameter in the correct analysis of neutron diffraction data, Eq.8. This is experimentally confirmed for the first two superlattice lines for order-disorder transition in the alloy beta-brass [27]. For the assumption that thermal vibrations can be ignored, the different errors for different lines *G* in Table 2 must be subtracted from the correct single-valued parameter in order to compare theoretical and experimental order parameters. This leads to multiple-valued experimental order parameters, one from each reciprocal line *G*. Alternately, the multiple values of the order parameter arise from the fact that it can be determined from *each superlattice line* by setting *W=0* in the correct expression, Eq.8. Crucially, these multiple values are equally valid since Eq.8 is valid for all lines *G*. This is a *conceptually very different* condition from the routine practice of neglecting physical phenomena that contribute negligibly to the problem under consideration. For example, ignoring DWF in Eq.3 leads to band gaps in semiconductors that are in error at finite temperatures but they are still single-valued. *To the best of our knowledge, this is the first report of a unique consequence, viz. an assumption of neglect of a physical phenomenon turns a single-valued experimentally* *observable quantity into a multiple-valued one and where all values are equally valid.* Therefore, this feature must be considered while judging the validity of the assumption (that thermal vibrations can be ignored) in addition to the usual feature of the magnitude of the errors caused by it. While it is acceptable to neglect physical phenomena that contribute negligibly as long as single-valued experimental observables remain single-valued, it is clearly unacceptable if it turns a single-valued experimental observable into a multiple-valued one. It is unacceptable even when the errors are relatively small e.g., Table2a, and can be acceptable only when errors are vanishingly small, in which case the multiple values become a single value for all practical purposes.

Thus, the only scientifically acceptable comparison is the one between theoretical models based on Eq.5 with diffraction data analyzed as per Eq.9a, i.e. both incorporating the role of thermal vibrations. Hence, *analysis of neutron diffraction data constrains the wavefunctions to be temperature dependent and must be represented as $\psi_i(T)$. It follows that the exchange interaction is also constrained to be temperature dependent, $J_{ij}(T)$, by neutron diffraction data.* This result is entirely consistent with high temperature electronic structure theory. We provide additional experimental and theoretical evidences that support the above conclusion.

An isotope effect due to zero-point vibrations is predicted to be a general feature [10] and has also been experimentally confirmed [28-30]. Recently, an isotope effect has been found [31] in an organic antiferromagnet that shifts $T_N$ (~ 2.5 K) by 4%. The authors attribute the lowering of $T_N$ to a reduction in the overlap of electronic wavefunction (or weaker exchange interaction) caused by smaller zero-point mean-square vibration amplitudes for the heavier isotope [31]. A similar explanation has also been given for the observed 0.6% isotope effect [32] in antiferromagnetic $La_2CuO_4$ ($T_N \sim 310$). The isotope effect is due to different zero-point vibrational amplitudes of different isotopes. As the temperature increases, the difference between vibrational amplitudes diminishes [10]. Therefore, one reason for the smaller isotope effect of 0.6% in $La_2CuO_4$ could be the high $T_N$ (~ 310 K).

From the above experimental results, it is clear that the observed isotope effect cannot be explained by models based on Eq.1. Also, the $T_N$ predicted from Eq.1 that ignores thermal vibrations is likely to be in serious error since the difference $J(\overline{u^2}) - J(0)$ is much larger than $J(\overline{u_1^2}) - J(\overline{u_2^2})$ that results in the 4% shift in $T_N$. Thus, only models based on Eq.5 can explain the above observed isotope effects.

In addition, microscopic theoretical models exist that suggest the same conclusion. For ferromagnetism in metals, a theoretical model for the temperature dependence of the effective exchange integral, $I_{eff}$, has been proposed [33] and the author concludes that "*at high temperatures $I_{eff}$ may be considerably smaller than at low temperatures.*"

In the case of insulators, a phonon mediated superexchange mechanism has been proposed [34-36]. This is a lattice dynamic effect and not due to Jahn-Teller interactions. An expression for the exchange intergral at finite temperatures has been derived as [34-36]

$$J(T) = J_0 - K_J T^6 \int_0^{\theta_D/T} x^5 (e^{-x} - 1)^{-1} dx \qquad (12)$$

where $J_0$ is the Heisenberg exchange at 0 K, i.e. $J_{ij}(0)$ in Eq.1, and $K_J$ is the spin-lattice coupling constant. Importantly, even at low temperatures, the zero-point contributions are found to be significant. This prediction is confirmed by the isotope effect in magnetism due to zero-point vibrations mentioned above.

Thus, the observed isotope effect [31,32] and the theory of phonon mediated superexchange [34-36] suggest that even when vibration amplitudes are relatively small at low temperatures, the exchange integral must be represented as temperature dependent, $J_{ij}(T)$. This is completely consistent with our conclusion based on the analysis of neutron diffraction data that *as a general principle, the exchange interaction must be represented to be temperature dependent, $J_{ij}(T)$, and Eq.5 must replace Eq.1 as the prototype model for magnetic phase transitions.*

In conclusion, we have addressed and resolved the fundamental contradiction that has existed from the earliest studies on magnetic phase transitions between theoretical models that ignore the role of thermal vibrations and represent the exchange interaction as a constant, $J_{ij}(0)$, and analysis of neutron diffraction data that always incorporates thermal vibrations even though it is also possible to analyze the same data by ignoring them. Ignoring thermal vibrations in both theoretical models and analysis of diffraction data leads to the latter giving different magnetic order parameters for different reciprocal lattice lines. *This appears to be the first report of a* *unique consequence, viz. the assumption to neglect a physical phenomenon turns a single-valued experimental observable into a multiple-valued one where all values are equally valid.* This assumption is clearly unacceptable and must be rejected. The second possibility of incorporating thermal vibrations in both leads to single-valued theoretical and experimental order parameters. Thus, analysis of neutron diffraction data constrain the exchange interaction in all theoretical models to be temperature dependent and represented as $J_{ij}(T)$. Additional experimental and theoretical evidences in support of this conclusion are presented.

**Table 1 Theoretical and experimental magnetic order parameters that must be used with different prototype Heisenberg models.**

| Theoretical model | Theoretical order parameter | Experimental order parameter | Comments |
|---|---|---|---|
| **Eq.1** – static lattice model $$H(0) = -\sum J_{ij}(0)\, \mathbf{S_i}.\mathbf{S_j}$$ | $\eta^{th}(T/T_c, 0)$ | $\eta_G^{exp}(T/T_c, 0)$ | $J_{ij}(0) \Leftrightarrow (\psi_i(0), \psi_j(0))$ $\Leftrightarrow W = 0$ in Eq.3 $\Rightarrow f = f(0)$ or $W = 0$ in Eq.6 and Eq.8 $\Rightarrow$ Eq.9b<br><br>Scientifically invalid as $\eta_G^{exp}(T/T_c, 0)$ dependent on **G**. |
| **Eq.5** – finite temperature model $$H(T) = -\sum J_{ij}(T)\, \mathbf{S_i}.\mathbf{S_j}$$ | $\eta^{th}(T/T_c, W_D)$ | $\eta_G^{exp}(T/T_c, W_D)$ | $J_{ij}(T) \Leftrightarrow (\psi_i(T), \psi_j(T))$ $\Leftrightarrow W = W_D$ in Eq.3 $\Rightarrow f = f(0)\, e^{-W}$ or $W = W_D$ in Eq.6 and Eq.8 $\Rightarrow$ Eq.9a<br><br>Scientifically valid as $\eta_G^{exp}(T/T_c, 0)$ independent of **G**. |

**Table 2 Percentage errors due to the neglect of thermal vibrations for various reciprocal lines at low temperatures and near $T_N$.**

**Table 2a)**

|  | $\varepsilon_{vib}^{\%}(111, T)$ | $\varepsilon_{vib}^{\%}(311, T)$ | $\varepsilon_{vib}^{\%}(331, T)$ | $\varepsilon_{vib}^{\%}(511, T)$ |
|---|---|---|---|---|
| MnO (15 K) | 0.2 % | 0.6 % | 1.1 % | 1.5 % |
| CoO (10 K) | 0.1 % | 0.4 % | 0.7 % | 1.1 % |
| *CoO (~$T_N$)* | *0.4%* | *1.5 %* | *2.5 %* | *3.6 %* |
| NiO (10 K) | 0.1 % | 0.5 % | 0.8 % | 1.10 % |
| *NiO (~$T_N$)* | *0.5 %* | *1.8 %* | *3.1 %* | *4.4 %* |

**Table 2b)**

|  | $\varepsilon_{vib}^{\%}(111, T)$ | $\varepsilon_{vib}^{\%}(222, T)$ | $\varepsilon_{vib}^{\%}(331, T)$ |
|---|---|---|---|
| NiFe$_2$O$_4$ (300 K) | 0.9 % | 3.4 % | 5.3 % |
| NiFe$_2$O$_4$ (850 K ~$T_N$) | 2.4 % | 9.3 % | 14.4 % |